\documentclass[aps,prx,twocolumn,floatfix,letterpaper,longbibliography]{revtex4-2}
\usepackage{graphicx}
\usepackage{bm,ulem}
\usepackage{amsmath,amsfonts}
\usepackage{amsbsy}
\usepackage[colorlinks=true,linkcolor=blue,urlcolor=blue,citecolor=blue,breaklinks=true]{hyperref}
\usepackage[utf8]{inputenc}
\DeclareUnicodeCharacter{03BC}{\mbox{$\mu$}}
\DeclareUnicodeCharacter{2009}{ }
\usepackage{color}
\usepackage{amssymb}
\usepackage{verbatim}
\usepackage[utf8]{inputenc}
\usepackage{colortbl}

\usepackage{xcolor}

\definecolor{imcolor}{rgb}{0.5,0.,0.5}

\newcommand{\ket}[1]{| #1 \rangle}

\newcommand{\matrixel}[3]{\langle #1 | #2 | #3 \rangle}

\newcommand{\beq}{\begin{equation}}
\newcommand{\eeq}{\end{equation}}

\newcommand{\bea}{\begin{eqnarray}}
\newcommand{\eea}{\end{eqnarray}}

\newcommand{\vs}[1]{\mathbf{#1}}
\newcommand{\abs}[1]{\left| #1 \right|}
\newcommand{\avg}[1]{\left\langle #1 \right\rangle}

\begin{document}

\title{Learning quantum ground states in the space of measurement outcomes}

\author{Kartiek Agarwal}
\affiliation{Material Science Division, Argonne National Laboratory, Lemont IL 60439}

\date{\today}

\begin{abstract}
We investigate variational learning of quantum many-body ground states directly in measurement space using autoregressive neural networks. In particular, we represent quantum states via probability distributions of outcomes over a symmetric informationally complete positive operator-valued measure (SIC-POVM). The probability distribution is encoded in the parameters of an autoregressive neural-network-based on gated recurrent units (GRUs). Ground states are obtained by gradient descent that updates the probability distribution to minimize the energy with respect to a given Hamiltonian, while enforcing positivity constraints that ensure that the distribution of measurement outcomes correspond to a physical quantum state. We analyze the role of constraint enforcement (hierarchy of positivity conditions), variety of neural network architectures (multiple layers, dilation, and modifications of input data) in determining the success of this approach. We benchmark our approach on one-dimensional transverse-field Ising model and the Heisenberg model, along with gapping fields, for system sizes up to $L=128$, illustrating its efficacy across a wide variety of models. 
\end{abstract}

\maketitle

\section{Introduction}

Variational representations of quantum many-body states based on machine learning architectures have emerged as a powerful framework for studying strongly correlated system~\cite{carleo2017solving,carrasquilla2019reconstructing,lange2024architectures,vicentini2022netket}. Training of such neural-network based quantum states have been successfully employed towards computing ground states~\cite{carleo2017solving,luo2019backflow,sharir2020deep,hibat2020recurrent,pfau2020ab,nomura2021dirac,Psiformer2023,Viteritti2023,Rende2024}, excited states and dynamics~\cite{Choo2019,Schmitt2020,hartmann2019neural,Vicentini2019,Donatella2023}, and interpreting experimentally obtained tomographic data~\cite{Torlai2018,koutny2022neural,czischek2022data,torlai2023quantum,wei2024neural}, across a broad variety of physical systems including frustrated spin systems~\cite{carleo2017solving,hibat2020recurrent,nomura2021dirac,Viteritti2023,viteritti2025transformer}, strongly interacting electronic systems~\cite{luo2019backflow,pfau2020ab,li2025attention,teng2025solving,chen2025neural}, nuclear systems~\cite{Lovato2022,Fore2025}, and quantum chemistry~\cite{hermann2020deep,Psiformer2023}. 

%and  using a wide variety of architectures including restricted Boltzmann machines, convolutional networks, recurrent neural networks, autoregressive transformers, neural backflow, hidden-fermion constructions, and self-attention wavefunction ansatzes

Most existing neural-network-based approaches represent the quantum state directly in terms of wavefunction amplitudes in a fixed basis. While these methods have achieved considerable success, several conceptual and practical challenges remain. First, unlike tensor-network approaches~\cite{Schollwock2011} such as matrix product states and density matrix renormalization group (DMRG), the representational biases and limitations of generic neural-network wavefunctions are still not fully understood~\cite{Deng2017,Glasser2018,Park2020,lange2024architectures,yang2024can}. Second, optimization in wavefunction space generally requires handling highly nonlinear energy landscapes and often relies on stochastic reconfiguration or natural-gradient methods whose computational cost scales unfavorably with the number of parameters in the architecture~\cite{Sorella2005,Park2020,Stokes2020,Rende2024,Chen2024}. Third, extending such approaches to mixed states, dissipative dynamics, and experimentally constrained measurement settings remains nontrivial~\cite{Schmitt2020,schmitt2022jvmc,Vicentini2019}.

An intriguing alternative approach is to represent quantum states directly in terms of distribution of measurement outcomes. Positive operator-valued measures~\cite{NielsenChuang} (POVMs) provide a natural framework for such a description. Given an informationally complete POVM, the quantum state may be reconstructed entirely from the probabilities of the different effects of the POVM. In this picture, the fundamental object learned by the neural network is not a \textit{complex} wavefunction amplitude, but rather a \textit{real} normalized probability distribution over measurement outcomes. Since the POVM outcome distribution corresponds directly to experimentally measurable quantities, it is naturally compatible with teacher-forcing and supervised learning protocols using experimental data~\cite{schmale2022efficient,quek2021adaptive}. Moreover, the same framework applies to both pure and mixed states, and dissipative dynamics. Finally, it is also well known that the choice of basis can strongly affect the ability of a neural network to learn a wavefunction even if it can represent it, providing further impetus to explore other ways of representing quantum states~\cite{yang2024can}.  

Despite these attractive features, learning ground states in POVM space also brings its own set of challenges. An important conceptual difference, as we discuss, comes from the fact that while every physical density matrix produces a normalized POVM outcome distribution, the converse is not true---arbitrary normalized distributions generally reconstruct to non-physical density matrices. This effectively converts the problem from one of purely minimizing the energy to the potentially more daunting task of minimization in the presence of constraints that enforce the positivity of the density matrix inferred from the POVM outcome distribution. 

This problem is closely related to the long-standing $N$-representability problem in reduced density matrix theory~\cite{Coleman1963,Garrod1964,Erdahl1987,Mazziotti2002,coleman2007reduced,Liu2007}. In particular, previous work by Mazziotti and others has shown that positivity conditions on appropriately constructed operator Gram matrices, $M_{ij} \equiv \mathrm{Tr} \left[ \rho O^\dagger_i O_j \right] \succeq 0$, provide powerful necessary conditions for physicality and representability~\cite{Mazziotti2002,Mazziotti2004,Mazziotti2012,Mazziotti2012Significant,Mazziotti2023}. Similar positivity conditions appear in broader semidefinite-programming hierarchies for quantum correlations and noncommutative polynomial optimization~\cite{Garrod1964,Navascues2008,Pironio2010,Baumgratz2012}. 

Another difference is that the complexity of the loss landscape is not determined by energy minimization, which becomes linear in the probabilities of POVM outcomes, but by the set of constraints which can be incorporated as a loss term that is nonlinear in the outcome probabilities. This also suggests that physically motivated constraints can be employed to better control the biases inherent in the training process, and refined as needed for the particular physical system at hand. 

A main aim of this work is to demonstrate that one can indeed obtain fairly accurate computation of the ground states working in the POVM basis. In the present work, we particularly focus on gated recurrent unit- (GRU-) based autoregressive neural-network architectures~\cite{hibat2020recurrent}, which are easy to sample and much less memory intensive to train, than for instance, large transformer-based models which feature all to all attention. For simplicity, we focus on one-dimensional gapped and gapless spin$-1/2$ models, but extensions to higher spins and higher dimensions is straightfoward and will be explore subsequently. We systematically analyze the effects of architectural choices, dependence on the weight and range of operators that enter into the momentum-resolved positivity constraints, the role of stochastic sampling noise, and adaptive balancing of constraint objectives.

Broadly, we find that the method can be adapted to obtain ground states of quantum systems with over hundred qubits, at least for the gapped and gapless one-dimensional spin models we have tested. We find that the ability of the model to reproduce long-range critical correlations, for instance, depends sensitively on both the recurrent architecture and the structure of the PSD constraints. For instance, we find that encoding physically motivated inductive biases, such as staggered parity information when considering antiferromagnetic critical models, dramatically improves the representation of staggered critical correlations. We find, perhaps surprisingly, that the spatial range of operators that enter the Gram matrices tends to play a rather weak role in the determination of constraints, even for gapless models, and increasing the range may even lead to noisier Gram matrices that hinder training. Practically, this means that we can restrict operator range to a finite number, which then implies that the computational complexity of the algorithm scales as $\sim \mathcal{O} (L^2)$, where $L$ is the number of qubits. 

While the algorithm tends to achieve similar accuracy for correlations in both gapped and gapless models, we find that the treatment of sampling noise in the gapless case is more subtle. In particular, the momentum-dependent constraint matrices $M^{(k)}_{ij}$ are noisy because they are built from expectation values that are sampled from the underlying POVM outcome distribution. One can estimate the effect of sampling noise on the eigenvalues of the Gram matrix by partitioning the samples into chunks and noting the typical differences in the eigenvalues, $\Delta \lambda$, between Gram matrices computed from chunks of the samples. In lieu of strict positivity, to account for noise, one demands that eigenvalues of $M^{(k)}_{ij} > - \tau_k = - \tau \Delta \lambda$ where $\tau$ is an $\mathcal{O} (1)$ number. For gapped models studied, we find that simply setting $\tau = 1.0$ provides reasonably accurate results that systematically improve upon increasing sampling batch sizes. For gapless models, the value of $\tau$ can be quite different from $1.0$. Crucially, it is independent of system size and the optimal value of $\tau$ can be inferred by comparing the correlations and energy of the trained state at small system sizes to that obtained from methods like DMRG or exact diagonalization. We are also able to directly relate the optimal value of $\tau$ to the actual distribution of near-zero eigenvalues of the Gram matrices obtained from sampling POVM outcomes with probabilities evaluated from the exact ground state for extremely small system sizes, $L = 8$. 

%A central observation of this work is that the optimization landscape is governed far more strongly by the positivity constraints than by the energy objective itself. In particular, we find that the ability of the model to reproduce long-range critical correlations depends sensitively on both the recurrent architecture and the structure of the PSD constraints. We also find that encoding physically motivated inductive biases, such as staggered parity information, dramatically improves the representation of certain critical correlations.

%At the same time, our results also reveal limitations of the approach. While local observables and energies can often be reproduced accurately, long-range critical correlations remain difficult to capture robustly. Increasing the complexity of the positivity constraints does not necessarily improve the results and can in fact destabilize optimization. These findings suggest that the representational and optimization challenges in POVM space differ qualitatively from those encountered in conventional neural quantum states.

The remainder of this work is organized as follows. In Sec.~\ref{sec:povmtheory}, we review the POVM representation, its connection to classical shadows, and the physicality constraints imposed on the inferred density matrices. In Sec.~\ref{sec:compare}, we situate our work in comparison to reduced density matrix approaches, and our previous shadow-tomography inspired approach; Ref.~\cite{rozon2025learning}. In Sec.~\ref{sec:architecture}, we introduce the autoregressive dual-stream recurrent architecture, momentum-resolved positivity constraints, and the associated training objective and training policy. In Sec.~\ref{sec:numerics}, we present numerical results for both critical and gapped one-dimensional spin models and analyze the dependence on architecture, operator range, and sampling noise. Finally, in Sec.~\ref{sec:conclusions}, we discuss the implications and limitations of POVM-space optimization and outline possible directions for future work.

\section{Representation in POVM Space and Physical Constraints}
\label{sec:povmtheory}

\subsection{Qubit POVMs, inversion, and statistical estimation}

A quantum measurement is described by a positive operator-valued measure (POVM), defined by a set of effects $\{E_i\}$ satisfying
\begin{equation}
E_i \ge 0, \qquad \sum_i E_i = \mathbb{I}.
\end{equation}
For a quantum state $\rho$, the probability of outcome $i$ is given by
\begin{equation}
p_i = \mathrm{Tr}(\rho E_i) \ge 0, \qquad \sum_i p_i = 1,
\end{equation}
which defines a normalized probability distribution.

If the effects span the operator space, the POVM is informationally complete and the complete state can be reconstructed as
\begin{equation}
\hat{\rho} = \sum_i p_i F_i,
\end{equation}
where $\{F_i\}$ is a dual operator frame.

\subsubsection{Tetrahedral SIC POVM} We now specialize to spin-$1/2$ which is the main interest in this work and where the effects and corresponding dual operators can be readily stated. Any qubit density matrix can be written as

\begin{equation}
\rho = \frac{1}{2}(\mathbb{I} + \mathbf r \cdot \boldsymbol\sigma),
\label{eq:dmqubit}
\end{equation}

\begin{figure}
\centering
    \includegraphics[width=0.5\columnwidth]{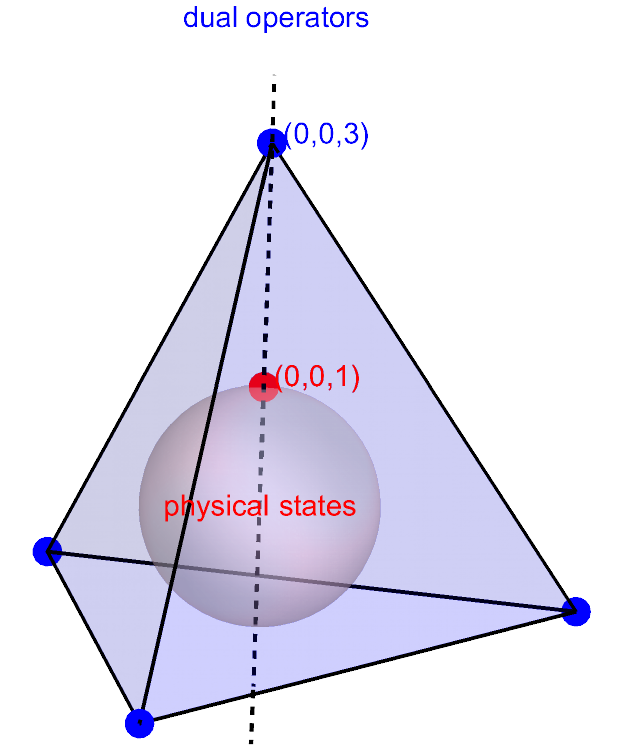}
    \caption{POVM dual operators $F_i$ (blue dots) for a single qubit sit at the ends of the minimal tetrahedron which fully encapsulates the Bloch (unit) sphere representing all possible physical states. The coordinates $(x,y,z)$ reflect the amplitude of $\sigma_x, \sigma_y, \sigma_z$ in the dual operators. The inferred density matrix resides in the convex hull of the tetrahedron by construction. It can capture all the possible physical states, but for certain values of the POVM outcome probabilties, one also can obtain unphysical states. Increasing the number of POVM outcomes does not result in the shrinking of this convex hull to more tightly wrap the space of physical states.}
    \label{fig:povm}
\end{figure}

where $\mathbf r$ is the Bloch vector with magnitude $r \le 1$, and $\boldsymbol{\sigma} = (\sigma_x, \sigma_y, \sigma_z)$ is the vector of Pauli operators.

A canonical example is the tetrahedral symmetric informationally complete (SIC) POVM with four effects
\begin{equation}
E_i = \frac{1}{4}(\mathbb{I} + \mathbf n_i \cdot \boldsymbol\sigma),
\end{equation}
where the vectors $\mathbf n_i$ form the vertices of a regular tetrahedron,
\begin{equation}
\sum_i \mathbf{n}_i = \vs{0}, \qquad \mathbf n_i \cdot \mathbf n_j = -\frac{1}{3}, \quad i \neq j. 
\end{equation}

Explicitly, we may choose, for instance

\begin{align}
    \mathbf{n}_1 &= \frac{1}{\sqrt{3}} \left[ 1, 1, 1 \right], \nonumber \\
    \mathbf{n}_2 &= \frac{1}{\sqrt{3}} \left[ 1, -1, -1 \right], \nonumber \\
    \mathbf{n}_3 &= \frac{1}{\sqrt{3}} \left[ -1, 1, -1 \right], \nonumber \\
    \mathbf{n}_4 &= \frac{1}{\sqrt{3}} \left[ -1, -1, 1 \right]. \nonumber \\
\end{align}

It is easy to confirm that in this case, the dual operators are given by
\begin{equation}
F_i = \frac{1}{2} \left( \mathbb{I} + 3 \mathbf n_i \cdot \boldsymbol\sigma \right). 
\end{equation}

Combined with the probabilities of the outcomes

\begin{align}
    p_i = \frac{1}{4} \left( 1 + \vs{r} \cdot \vs{n}_i \right), 
\end{align}

we recover the density matrix associated with the probability distribution $p_i$ as $\hat{\rho} = \sum_i p_i F_i$. 

\subsubsection{Non-physicality of dual operators}

It is worth noting that the dual frame operators $F_i$ do not correspond to physical density matrices. Explicitly, we can see this by noting the form of physical single-qubit density matrices, as seen in Eq.~(\ref{eq:dmqubit}), characterized by the vectors $\mathbf{r}$ with magnitude $r \le 1$. The operator $F_i$ are of this form with $r = 3$, implying a density matrix with an unphysical length of the Bloch vector. 

Readers familiar with classical shadow tomography~\cite{Huang2020} may also recognize this factor of $3$ which arises in the standard inverse shadow channel---the operator $F_i$ is precisely the inferred or estimated density matrix $\hat{\rho}$ corresponding to a single shadow tomography measurement outcome on a single qubit. In particular, for the random measurement axis defined by the Bloch vector $ \mathbf{u}_i$ and measurement outcome  $b_i = \pm 1$, the corresponding effect vector is given by $\mathbf{n}_i = b_i \mathbf{u}_i$, and the estimated density matrix $\hat{\rho} = F_i$. 

The non-physicality of the $F_i$ has important implications for the density matrix inferred as $\hat{\rho} = \sum_i p_i F_i$. In particular, it is easy to check that $\text{Tr} \left[ \hat{\rho} \right] = 1$, that is, the density matrix is normalized. However, even though the probability distribution defined by the outcome probabilities $\{ p_i \}$ is a legitimate normalized probability distribution, the inferred density matrix $\hat{\rho}$ may not be a legitimate density matrix. A simple example is obtained by setting any one $p_i$, say $p_1 = 1$ and the rest zero---the eigenvalues of the inferred density matrix then are $3/2, -1/2$ which sum to $1$ but clearly do not represent a physical state of a qubit. Thus, extra conditions need to be imposed on the distribution $\{p_i\}$ to ensure the density matrix is positive semi-definite (PSD), that is, $\rho \succeq 0$.  

We note that there is a simple geometric reason for the non-physicality of $F_i$. Given $p_i>0, \sum_i p_i = 1$, the estimated density matrix $\hat{\rho} = p_i F_i$ lies in the convex hull of the matrices $F_i$. For informational completeness, this convex hull must at least encompass the entire Bloch sphere. For the four effects and corresponding dual operators of the tetrahedral POVM, this can clearly only be achieved by making the $F_i$ correspond to operators that lie outside the Bloch sphere; see Fig.~\ref{fig:povm}. 

\subsubsection{General qubit POVM}

Given the above discussion, it may be tempting to consider increased number of POVM effects and dual operators in the hope that the convex hull of these dual operators now tightly wrap around the Bloch sphere. Unfortunately, this is not the case, and increasing the number of effects does not change the factor of $3$ in the $F_i$. We note that this result is evident from the analogous classical shadows derivations; one can go to the limit of infinite effects (Haar random measurement bases), and the inverted density matrix per sample continues to be unphysical with the same magnitude of the Bloch vector, $r = 3$. 

We can see this explicitly as follows. A general qubit POVM can be expressed in terms of its effects

\begin{equation}
E_i = w_i (\mathbb{I} + \mathbf n_i \cdot \boldsymbol\sigma),
\label{eq:Ei}
\end{equation}

where $\mathbf n_i$ are unit Bloch vectors and $w_i > 0$ satisfy

\begin{equation}
\sum_i w_i = 1, \qquad \sum_i w_i \mathbf {n}_i = \mathbf{0}.
\end{equation}

The above conditions ensure the positivity conditions on the effects $E_i$, and their normalization $\sum_i E_i = \mathbb{I}$. 

The probabilities of the outcomes $i$ can then be computed using Eqs.~(\ref{eq:dmqubit},\ref{eq:Ei})

\begin{equation}
p_i = w_i (1 + \mathbf r \cdot \mathbf n_i).
\end{equation}

If the measurement directions form an isotropic spherical two-design,
\begin{equation}
\sum_i w_i n_i^\alpha n_i^\beta = \frac{1}{3} \delta_{\alpha\beta},
\end{equation}
then
\begin{equation}
\sum_i p_i \mathbf n_i = \frac{1}{3} \mathbf r,
\end{equation}
which implies
\begin{equation}
\mathbf r = 3 \sum_i p_i \mathbf n_i, \qquad F_i = \frac{1}{2} \left(1 + 3 \mathbf{n}_i \cdot \mathbf{\sigma} \right).
\end{equation}

This establishes the universality of the factor of three, independent of the number of POVM effects, provided the directions form a two-design. 

The reason for the unphysicality of the dual operators $F_i$ despite the increased effect count is fundamentally due to the non-commutative nature of the effects, which results in subtle geometric constraints on the probabilities $p_i$, if they are obtained from a true physical state $\rho$. 

\subsubsection{Finite sample variance}

Finally, it is important to note that, to find ground states of quantum systems comprising hundreds of qubits, we will need to sample the distribution $\{p_i\}$. It is therefore worth understanding the sampling variance of various observables. 

Let $A$ be a single qubit observable with decomposition
\begin{equation}
A = \mathbf{a} \cdot \boldsymbol{\sigma}, \qquad \abs{\mathbf{a}} = 1.
\end{equation}

We have chosen $A$ to be traceless without loss of generality as the part proportional to the identity merely shifts the mean expectation value and does not affect the variance.

Using the dual frame representation $\hat{\rho} = \sum_i p_i F_i$, the expectation value can be written as
\begin{equation}
\langle A \rangle = \mathrm{Tr}(\rho A) = \sum_i p_i \mathrm{Tr}(F_i A), 
\end{equation}

which implies an unbiased estimator of the form
\begin{equation}
\hat{A} = \frac{1}{N} \sum_{k=1}^N A_{i_k}, \qquad A_i := \mathrm{Tr}(F_i A),
\end{equation}
where $i_k$ are sampled according to the probability distribution $\{p_i\}$. The variance can then be computed as
\begin{equation}
\mathrm{Var}(\hat{A}) = \frac{1}{N} \left( \sum_i p_i A_i^2 - \langle A \rangle^2 \right) \le \frac{1}{N} \sum_i p_i A^2_i.
\end{equation}

Using the explicit form of the dual operators, we find $A_i = 3 \mathbf{a} \cdot \mathbf{n}_i$. Thus, $A^2_i \le 9 \abs{\vs{a}}^2 = 9$. And since the probabilities $\{p_i\}$ constitute a normalized probability distribution, we find 

\begin{align}
    \mathrm{Var}(\hat{A}) \le \frac{9}{N}.
    \label{eq:var}
\end{align}

We note that this result is precisely that from classical shadow tomography. This result merely follows from the boundedness of the norm of the operator $A$ and the two-design property of the POVM effects. 

\subsubsection{Multiple qubits}

We will assume that the POVM effects over $n$-qubits are simply the product of the effects on each site---

\begin{align}
    E_{i_1,...i_n} = E_{i_1} \otimes \dots \otimes E_{i_n}. 
\end{align}

The dual frame operators follow analogously to the single qubit case and are merely products of the individual site operators. Thus, 

\begin{align}
    F_{i_1,...i_n} = F_{i_1} \otimes \dots \otimes F_{i_n}. 
\end{align}

Unbiased estimators of an observable $A$ can now we extracted by sample averaging $A_{\vs{i}} := \mathrm{Tr} \left[ F_{\vs{i}} A \right]$. The unbiased estimator of an $n$-qubit weight-$w$ operator of the form 

\begin{align}
A = \prod_{j=1}^{j=w} \vs{a}_{i_j} \cdot \vs{\sigma}_{i_j}
\end{align}

is given by

\begin{align}
\hat{A} = \frac{1}{N} \sum_{k=1}^{N} A_{\vs{i}_k}, \qquad A_{\vs{i}} = \mathrm{Tr} \left[ F_{\vs{i}} A \right].
\end{align}

where 

\begin{align}
    \abs{A_{\vs{i}}} \le 3^w, 
\end{align}

and $w$ is the weight of the operator $A$. The variance of such an operator is then bounded analogously as before with

\begin{align}
    \textrm{Var}{[\hat{A}]} \le \frac{3^{2w}}{N}. 
    \label{eq:var_w}
\end{align}

The result Eq.~(\ref{eq:var_w}) informs us that low-weight observables are easier to estimate, requiring fewer samples, than large weight observables. This result follows whether or not the probability distribution $\{p_i\}$ describes a physical density matrix. This has important consequences for the kind of constraints we will impose to ensure that the learned distribution of outcomes probabilities $\{p_i\}$ correspond to physical states, which we turn to next.  

Finally, it worth comparing the sampling variance in the present framework to that in wavefunction based methods where one instead draws samples from the squared wavefunction amplitude $\left| \Psi(\sigma_1, \sigma_2, \cdots) \right|^2$ to estimate observables. For instance, for the energy, one has 

\begin{align}
\left\langle E \right \rangle &=\sum_\mathbf{\sigma} p_\vs{\sigma} E_{\text{loc}} (\vs{\sigma}), \nonumber \\
E_{\text{loc}} (\mathbf{\sigma}) &= \sum_{\sigma'} \matrixel{\sigma}{H}{\sigma'} \frac{\Psi_{\sigma'}}{\Psi_{\sigma}}
\end{align}

where $E_{\text{loc}}$ serves as a local estimator that can be averaged over all configurations with appropriate probabilities to arrive at an estimate of the energy of the state. Clearly in this case, if the wavefunction $\Psi$ has nodes, the sampling variance of $E_{\text{loc}}$ can get arbitrarily large. Nodes may arise due to symmetry, Pauli exclusion in fermionic systems, and frustration in spin models. A nice aspect of the POVM framework is that the variance of local observables is strictly bounded from above.  

%which implies concentration bounds such as
%\begin{equation}
%\Pr(|\overline O_N - \mathbb E[\widehat O]| > \epsilon)
%\le 2 \exp\left(-\frac{2N\epsilon^2}{(6|\mathbf o|)^2}\right).
%\end{equation}

\subsection{Physical constraints on the probability distribution}

Our goal is to find the distribution $\{p_{\vs{i}}\}$ which minimizes the energy of the inferred state $\hat{\rho}$ with respect to the target Hamiltonian $H = \sum_j h_j P_j$, where $P_j$ are Pauli operators which form a convenient basis to represent all Hermitian observables that may enter the Hamiltonian $H$. In particular, the expectation value $\langle \hat{H} \rangle$

\begin{align}
    \langle \hat{H} \rangle = \sum_j h_j \langle \hat{P}_j \rangle = \sum_{\vs{i},j} h_j  p_{\vs{i}} \mathrm{Tr} \left[ F_{\vs{i}} P_j \right], 
\end{align}

is purely linear in the probabilities $p_{\vs{i}}$ which can be trivially optimized. Of course, this would yield incorrect results in general since the optimized distribution $p_{\vs{i}}$ will not necessarily correspond to a physical state, $\rho \succeq 0$. 

In this protocol, geometry of the optimization landscape thus comes almost entirely from the constraints imposed on the distribution $p_{\vs{i}}$ from demanding the physicality of the underlying state $\rho$. This is in contrast to the setting where one directly learns the complex wavefunction amplitudes in a fixed `computational basis'. The energy itself provides a quadratic landscape in terms of the complex amplitudes $\psi_{\vs{i}}$. 
%Of course, internally, the neural network may have multiple heads that independently predict 

\subsubsection{Single qubit case}
%While any normalized $\{p_\vs{i}\}$ defines an operator $\rho = \sum_i p_i F_i$, physical states require $\rho \succeq 0$. 

For a single qubit, once the trace of the density matrix is fixed to $1$ as is the case for the inferred $\hat{\rho}$, positivity only requires one further condition, equivalent to
\begin{equation}
|\mathbf r|^2 \le 1,
\end{equation}
or
\begin{equation}
\left| 3 \sum_i p_i \mathbf n_i \right|^2 \le 1.
\end{equation}

Thus, for a single qubit, one must merely ensure a single quadratic constraint on the probabilities $\{p_i\}$ along with attempting to minimize $\avg{\hat{H}}$ which is linear in the probabilities. Given that neural networks more stably output such probabilities from task heads, as opposed to complex wavefunction amplitudes, one may expect that the optimization problem in POVM outcomes may be more efficiently encoded in a variational neural network algorithm. This facet partly inspires the exploration of representing quantum states in the informationally complete POVM basis in this work. 

\subsubsection{Multiple qubits: Operator Gram matrices}

The positivity constraints for a density matrix on multiple qubits are considerably more involved. In general, given the fact that the density matrix scales exponentially in the number of qubits, this task may be expected to be exponentially hard. Indeed, prior work shows that related quantum representability problems are computationally hard~\cite{Liu2007}. 

To alleviate this issue, one can consider imposing conditions on reduced density matrices on subsystems of few qubits. A more robust approach is to consider reduced density matrices built out of expectation values of low-weight operators at all sites. This approach has been explored particularly in the context of quantum chemistry~\cite{Coleman1963,Mazziotti2012} where the Hamiltonian is generally given by at most two-particle interaction terms, and relevant observables are the one-particle and two-particle operators as well. A natural approach outlined by Mazziotti and others is to consider the space of all possible few-body operators (including non-Hermitian terms) $\{ O_i \}$ and construct a matrix 

\begin{align}
    M_{ij} = \mathrm{Tr} \left( \hat{\rho} O_i^\dagger O_j \right)
\end{align}

which is then enforced to be PSD, that is, $M \succeq 0$. Enforcing the positivity of $M$ is equivalent to demanding that all operators $A = \sum_i c_i O_i$ satisfy the condition $\mathrm{Tr}(\rho A^\dagger A ) \ge 0$. 

(We note here that one should be careful to distinguish this matrix from a covariance matrix, such as those relevant for computing the quantum Fisher information at zero temperature. Covariance matrices tend to constraint the density matrix more weakly as they check for the positivity of $\mathrm{Tr} (\hat{\rho} B^\dagger B) $ for $B = \sum_i c_i \tilde{O}_i$ and $\tilde{O}_i = O_i - \mathrm{Tr}(\hat{\rho} O_i)$ is the centered version of $O_i$.) 

The natural extension of these ideas in the qubit setting is to work with Pauli operators of maximum fixed weight. In this work, our main physicality constraint will be the enforcement of the positivity of $M_{ij}$ formed by computing overlaps of all (Hermitian) Pauli observables $O_i$ of maximum fixed weight $w = 2$ and range $R$. Here, the range $R$ is the maximum distance between the sites at which the operator $O_i$ acts non-trivially. We note that constraining expectation values of low weight observables is also natural in our setting because the variance of higher weight observables grows exponentially in the weight, making the sampling of such observables with sufficient accuracy untenable in practice. 

\subsubsection{Variance and Purity Constraints}

Besides these constraints, one can also enforce constraints on the variance and the purity of the state. The variance constraint is given by 

\begin{align}
    \mathrm{Var}[H] = \mathrm{Tr}\left[\hat{\rho} H^2 \right] - \mathrm{Tr}\left[\hat{\rho} H \right]^2 = 0,  
\end{align}

for states $\hat{\rho}$ corresponding to the eigenstates, including the ground state, of the many-body system. In practice, we find that this constraint does not help in training as it tends to not differentiate between eigenstates. We thus do not employ it for the results presented in this manuscript.  

Another constraint that works similarly to the variance constraint but is more general, and which can be implemented fairly efficiently is to demand that the purity of the inferred density matrix is 1, that is, 

\begin{equation}
\mathrm{Tr}(\hat{\rho}^2) = \sum_{\vs{i} \vs{j}} p_\vs{i} p_\vs{j} \mathrm{Tr}(F_\vs{i} F_\vs{j}) = 1. 
\end{equation}

This constraint is explicitly quadratic in $p_i$ for any number of qubits. For the tetrahedral SIC POVM considered in this work, $\mathrm{Tr} (F_i F_j) = 6 \delta_{ij} - 1$ at the single qubit level. Due to this condition, the contribution of terms proportional to $p_\vs{i} p_\vs{j}$ where the outcomes differ at $x$ sites decays in magnitude as $5^x$, faster than the number of such terms which grows as $3^x$. Thus, we can expect that the purity is dominated by terms involving $p_\vs{i} p_\vs{j}$ where the respective outcomes $\vs{i}, \vs{j}$, differ only at a few sites. For simplicity, in this work, we have not tested whether such a purity constraint may help training, focusing on the positivity constraints obtained by directly probing the matrix $M_{ij}$ discussed above. 

\section{Relation to previous reduced density matrix and shadow-based approaches}
\label{sec:compare}

The present work is conceptually related both to reduced density matrix (RDM) approaches based on positivity constraints~\cite{Mazziotti2002,Mazziotti2012} and to our previous work on optimizing finite ensembles of classical shadows~\cite{rozon2025learning}. However, the present formulation differs from both in important ways.

\subsection{Comparison to reduced density matrix methods}

In the reduced density matrix framework pioneered by Mazziotti and others, one considers expectation values of a collection of operators $\{O_i\}$ as independent variational degrees of freedom. The positivity constraints are then enforced by demanding positivity of Gram matrices of the form

\begin{align}
M_{ij}
=
\mathrm{Tr}\left(
\rho O_i^\dagger O_j
\right).
\end{align}

In this perspective, the correlators entering the matrix $M$ are treated as unconstrained optimization variables.

In contrast, in the present work, the correlators are not optimized independently. Instead, all expectation values arise from a single underlying POVM outcome distribution $\{p_{\vs{i}}\}$ generated autoregressively by the neural network. Consequently, the correlators entering the Gram matrix inherit nontrivial geometric constraints already at the level of the POVM representation itself.

This distinction is particularly transparent at the single-qubit level. For the tetrahedral SIC POVM considered here, the reconstructed Bloch vector satisfies

\begin{align}
\vs{r}
=
3
\sum_i p_i \vs{n}_i,
\end{align}

with the probabilities constrained to lie within the simplex $p_i \ge 0$, $\sum_i p_i = 1$. Although the dual operators $F_i$ themselves are unphysical, the resulting expectation values cannot vary arbitrarily. Increasing the magnitude of one spin component necessarily restricts the accessible values of the others due to the geometry of the tetrahedral frame. For example, large $\langle Z \rangle$ suppresses the simultaneously accessible values of $\langle X \rangle$ and $\langle Y \rangle$.

Thus, unlike standard reduced density matrix optimization where correlators may independently drift into strongly incompatible regions before being corrected by PSD constraints, the POVM representation already builds in a degree of compatibility between observables through the underlying probability simplex. One may therefore view the present approach as introducing a structured parametrization of the correlator space prior to imposing the higher-body positivity constraints.

At the same time, the POVM representation introduces new challenges absent in conventional RDM methods. In particular, observables are estimated stochastically from sampled POVM outcomes, leading to finite-sampling fluctuations in the Gram matrices. These fluctuations become increasingly important for higher-weight operators and near-null modes of the Gram matrix, which show some qualitative differences between gapped and gapless models; see discussion in Sec.~\ref{sec:gappedvsgapless}. 
 
\subsection{Comparison to optimized shadow ensembles}

The present work is also closely related to our earlier work~\cite{rozon2025learning}, where finite ensembles of classical-shadow outcomes were directly optimized variationally.

In that approach, one considers a fixed collection of shadow outcomes,
\begin{align}
\left\{
\left( \mathcal{U}^{(1)}, b^{(1)} (\theta) \right)
\dots,
\left( \mathcal{U}^{(N)}, b^{(N)} (\theta) \right)
\right\},
\end{align}
and optimizes the discrete measurement outcomes $b^{(i)}$ in order to minimize the energy. In particular, the Haar-random unitaries $\mathcal{U}^{(i)}$ for each measurement sample $i$ fixes the basis of measurement and are independent of model parameters, while the discrete measurement outcome $b^{(i)} = \ket{0 1 \cdots 1}$ etc. are governed by model parameters and optimized. Since each shadow outcome contains local measurement information on every site, the number of optimized variables scales extensively as
\begin{align}
\mathcal{O}(LN),
\end{align}
where $L$ is the system size and $N$ is the ensemble size.

By contrast, in the present work, the neural network parametrizes the full probability distribution autoregressively. The total number of trainable parameters is therefore independent of system size once the architecture is fixed. Increasing the system size merely corresponds to unrolling the same recurrent architecture over longer sequences.

Conceptually, the previous shadow-ensemble approach may be viewed as directly optimizing a finite empirical distribution over measurement outcomes, whereas the present work instead learns a generative model capable of sampling from the full distribution. This improves scalability and allows the same architecture to be applied to large systems without increasing the parameter count. It is likely however that the optimization landscape in the present work differs from that obtained in Ref.~\cite{rozon2025learning} given that the measurement sample used to infer the state and correlators is largely unchanged between training steps in that work while in this work, each training step invokes an entirely new sample of outcomes from the underlying POVM distribution. This potentially makes optimization harder, but more robust. 

\section{Model Architecture, Loss Function, and Training Strategy}
\label{sec:architecture}

\subsection{Autoregressive model and dual-stream architecture}

We choose to model the distribution over POVM outcomes via an auto-regressive neural network. The auto-regressive choice enables efficient sampling of the underlying probability distribution stored in the neural network architecture. In general, autoregressive models have been applied with considerable success for both one-dimensional and two-dimensional quantum systems~\cite{sharir2020deep,hibat2020recurrent,nomura2021dirac,Bohrdt2024,Donatella2023}, with the consensus being that these models are more limited in the latter case. In this work, for simplicity, we restrict our attention to studying one-dimensional models using POVM outcome distributions and understanding how different physicality constraints and architectural changes affect the fidelity of learned wavefunctions. 
%In general, we anticipate transformer-based architectures with Monte-Carlo sampling would be more capable, particularly for computing ground states of higher dimensional systems.     

In particular, the POVM outcome probability for the $j^\text{th}$ site, $i_j \in \{0,1,2,3\}$ is given by 
\begin{equation}
p_{\vs{i}} = \prod_{j=1}^{L} p(i_j \mid i_{<j}),
\end{equation}
where $L$ is the system size. 

\begin{figure}
\centering
    \includegraphics[width=\columnwidth]{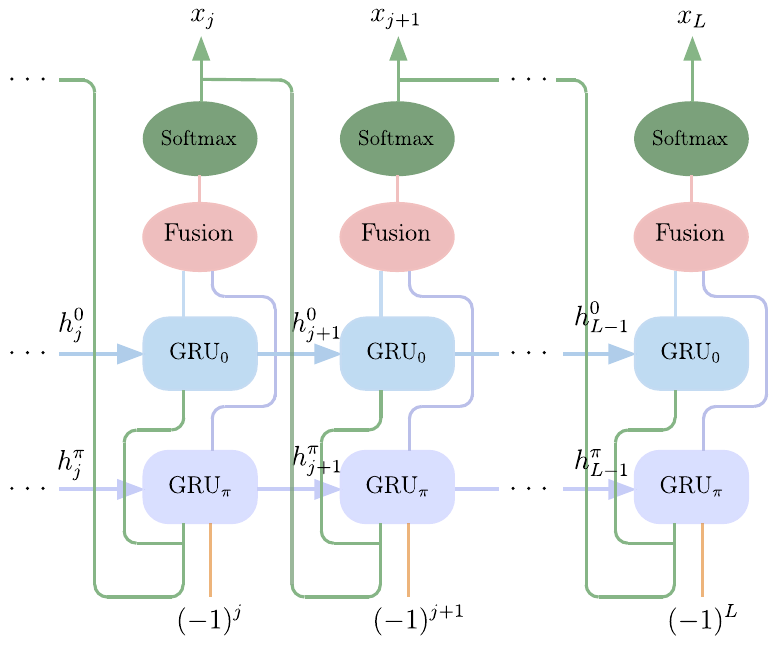}
    \caption{Dual stream architecture. Two parallel RNNs are simulated, $\text{GRU}_\pi$ which is fed the site parity $(-1)^j$ and, $\text{GRU}_0$ which is not. We can consider the two streams as being attuned to momentum $k = \pi, 0$ sectors, respectively. The output of these two GRU streams is fed into a Fusion head (details in main text), from which logits denoting the POVM measurement outcomes is output. These logits are finally used to compute outcome probabilities using a Softmax head.}
    \label{fig:GRU}
\end{figure}

The conditional probabilities are parameterized using gated recurrent units (GRUs). At each site $j$, a hidden state $h_j$ is updated sequentially,
\begin{equation}
h_j = \mathrm{GRU}(x_j, h_{j-1}),
\end{equation}
where $x_j$ encodes the previous outcome $i_{j-1}$ in a one-hot encoding.

\subsubsection{Dual-stream architecture.}

In general, we find that to capture long-range correlations, particularly in models with anti-ferromagnetic correlations, it becomes important to encode the parity of the site as an input into the GRU. We thus also consider architectures in which two independent GRU chains are maintained---

\begin{align}
h_j^{(0)} &= \mathrm{GRU}_0(x_j, h_{j-1}^{(0)}), \\
h_j^{(\pi)} &= \mathrm{GRU}_\pi([x_j, (-1)^j], h_{j-1}^{(\pi)}),
\end{align}

where the second stream receives the site parity $(-1)^j$ as an additional input.

The two streams are combined through a learned gate rather than by simple concatenation.  Denoting the final-layer hidden states by \(h_j^{(0)}\) and \(h_j^{(\pi)}\), we define\begin{equation}g_j = \sigma(W_g h_j^{(0)} + b_g),\end{equation}and construct the combined hidden representation\begin{equation}h_j^{\rm out}=h_j^{(0)}+g_j \odot h_j^{(\pi)}.\end{equation}

The role of the gate is to allow the staggered stream to act as a controlled correction to the uniform stream.  This is found to be important in antiferromagnetic critical systems, where the relevant long-distance correlations have a strong \(k=\pi\) component.  A single parity-aware stream tends to overemphasize staggered correlations and degrade uniform channels, while the gated dual-stream architecture allows the model to retain uniform structure while selectively activating the parity-dependent component.

The logits for the next POVM outcome are then obtained from a linear output layer,\begin{equation}\ell_j = W_{\rm out} h_j^{\rm out} + b_{\rm out},\end{equation}followed by a softmax,\begin{equation}p(i_{j+1}=a \mid i_{\le j})=\frac{\exp(\ell_{j,a})}{\sum_b \exp(\ell_{j,b})}.\end{equation} 

The main units of the dual-stream architecture are also illustrated in Fig.~\ref{fig:GRU}. 

\subsubsection{Stacked and dilated recurrent architectures}In addition to increasing the hidden dimension, a natural way to enhance the expressivity of recurrent models is to increase the number of stacked recurrent layers. In this setting, each layer maintains its own hidden state, and the input to layer $\ell$ at site $j$ is the hidden state from the previous layer at the same site,\begin{equation}h_j^{(\ell)} = \mathrm{GRU}^{(\ell)}\!\left(h_j^{(\ell-1)},\, h_{j-1}^{(\ell)}\right),\end{equation}with $h_j^{(0)} \equiv x_j$. The total depth of the model is controlled by $N_{\rm Layers}$, which is shared across both streams in the dual-stream architecture described above. Each layer also possesses its own learned initial hidden state.

Stacking multiple layers in this manner increases the nonlinear processing capacity of the model at each site, but does not significantly alter the effective range over which correlations can be propagated along the sequence. In practice, we observe that increasing $N_{\rm layers}$ beyond one or two layers does not substantially improve the learned long-range correlations in the systems considered here. 

We note that several works have also proposed using dilated recurrent neural networks to improve correlations obtained, particularly in the case of critical models which exhibit characteristic scale-free correlations~\cite{ayub2026geometry,Chang2017}. In a dilated GRU, the hidden state at site $j$ in layer $\ell$ depends not on the immediately preceding hidden state $h_{j-1}^{(\ell)}$, but on a hidden state separated by a dilation factor $d_\ell$,\begin{equation}h_j^{(\ell)} = \mathrm{GRU}^{(\ell)}\!\left(h_j^{(\ell-1)},\, h_{j-d_\ell}^{(\ell)}\right).\end{equation}By choosing dilation factors that grow with layer index, for example $d_\ell = 2^{\ell-1}$, one obtains an exponentially increasing receptive field with depth, allowing information to propagate over long distances with relatively few layers.

We have not experimented extensively with such dilated networks, though we find that the results do not improve significantly, at least up to $N_L = 2$. (Recent work however suggests that going to $N_L = 8$ may in fact deliver improvements at least for the stoquastic transverse-field Ising model; see Ref.~\cite{ayub2026geometry}.) All results presented in this work are therefore obtained using the stacked dual-stream GRU architecture with $N_L = 2$.

%At the boundaries where $j-d_\ell \le 0$, we adopt a simple padding strategy in which the corresponding hidden states are replaced by the learned initial state of that layer. This preserves translational structure away from the edges while avoiding the need for special boundary conditions.

\subsection{Training objective}

\subsubsection{Energy objective}

The expectation value of the Hamiltonian,
\begin{equation}
\langle \hat{H} \rangle = \sum_{\vs{i}} p_{\vs{i}} \sum_j h_j \mathrm{Tr}\left[F_{\vs{i}} P_j \right],
\end{equation}
is linear in the probabilities $p_{\vs{i}}$. The corresponding loss is
\begin{equation}
\mathcal{L}_E = \langle \hat{H} \rangle.
\end{equation}

As discussed previously, minimizing $\mathcal{L}_E$ alone is insufficient, as it does not enforce physicality of the inferred state.

\subsubsection{Momentum-resolved PSD constraints}

To enforce physicality, we impose positivity constraints on operator Gram matrices constructed from low-weight operators. In this work, we focus on translationally invariant models with periodic boundary conditions. To get clear signals from the PSD constraints, and to reduce the dimensionality of the Gram matrix to enable faster computation, it is useful to then construct the Gram matrices in momentum space. To this end, we define a set of template Pauli operators, $\{ O_i \}$, which are restricted by weight and range. For instance, for maximum weight $w = 2$ and range $R = 2$, the space of templates is $3 + 3 \cdot 3 = 12$ dimensional, consisting of $3$ single site Pauli operators and $9$ two site Pauli operators. Likewise, for maximum weight $w = 2$, and range $R = 4$, the dimension of template space is $30$, corresponding again to $3$ single site Pauli operators and $3 \cdot 3 \cdot 3$ two-site Pauli operators at maximum separation of $4$ sites. We restrict ourselves to maximum weight $w = 2$ and range $R = 8$. 

We then define a set of operators $\{O_i^{(k)}\}$ in each momentum sector $k$, obtained by Fourier transforming the template operators. 

The corresponding Gram matrix is
\begin{equation}
M^{(k)}_{ij} = \mathrm{Tr}(\hat{\rho} O_i^{(k)\dagger} O_j^{(k)}),
\end{equation}
which must satisfy
\begin{equation}
M^{(k)} \succeq 0.
\end{equation}

Violations of this condition are detected via the eigenvalues $\lambda^{(k)}_\alpha$ of $M^{(k)}$. 

\subsubsection{Handling stochastic noise via buffer and gradient batches}

In general, the Gram matrices are constructed by computing expectation values of operators by sampling the underlying POVM outcome distribution encoded in the RNN architecture. As a result, these expectation values are generally noisy approximations to the true expectation value of these operators and this stochasticity must be handled with care.  To reduce stochasticity, large batch sizes are preferable. However, this quickly becomes highly memory intensive if all the samples carry information about gradients needed for backpropagation. Thus, to reduce stochastic noise while allowing for gradients for training, we consider a two-step approach that breaks up the computation into a buffer step in which samples are generated in inference mode, and a gradient step in which all expectation values carry gradients for optimizing the neural network. 

In particular, during the buffer step, we compute expectation values of operators that enter the Gram matrices $M^{(k)}_{ij}$ by sampling the neural network in inference mode with a large batch of size $B$. This allows for larger buffer batch sizes $B$. (In practice, most of our results are computed for $B = 128,000$. Smaller buffer batch sizes generally produce noisier results while larger buffer batch sizes give diminishing returns.) The samples are further divided into two equal `training' and `validation' halves, with the `training' half used to compute the Gram matrices, $M^{(k)}_{\text{tr}}$, which are subsequently diagonalized to obtain eigenvalues $\lambda^{(k)}_{\alpha, \text{tr}}$ and eigenvectors $v^{(k)}_{\alpha,\text{tr}}$. 

The other `validation' half is used to construct Gram matrices $M^{(k)}_{\text{val}}$ which in general differ from $M^{(k)}_{\text{tr}}$ due to sampling fluctuations. The eigenvalues used to determine the violation of the PSD constraints are inferred as 

\begin{align}
    \lambda^{(k)}_{\alpha, \text{val}} = \left[ v^{(k)}_{\alpha,\text{tr}} \right]^\dagger \cdot M^{(k)}_{\text{val}} \cdot v^{(k)}_{\alpha,\text{tr}}, 
\end{align}

that is, from a combination of the training and validation set of samples. 

The reason for adopting this approach is two-fold. First, diagonalizing the stochastic Gram matrices tends to produce overly negative eigenvalues that benefit from directions in which noise can add up constructively. Computing the eigenvalues $\lambda^{(k)}_{\text{val}}$ using the eigenvectors from the tends to reduce this effect. Second, the differences between eigenvalues,  

\begin{equation}
\Delta \lambda^{(k)}_\alpha = |\lambda^{(k)}_{\alpha,\mathrm{tr}} - \lambda^{(k)}_{\alpha,\mathrm{val}}|.
\end{equation}

provides an unbiased estimation of noise in each momentum, denoted by $k$, sector, due to stochastic fluctuations. While the true Gram matrices can be expected to be fully PSD, the Gram matrices computed stochastically may have negative eigenvalues even when the underlying POVM outcome distribution corresponds to a physical state. Thus, it is important to set tolerances for the eigenvalues $\lambda^{(k)}_{\text{val}}$ based on some measure of the effect of stochasticity due to finite sampling noise. 

\subsubsection{PSD constraint objective}

We can now define the PSD constraint objective. This is given by

\begin{align}
    \mathcal{L}_{\mathrm{PSD}} = \sum_{k,\alpha} f (\lambda^{(k)}_{\alpha, \text{val}}, \tau_k, s_k ) \; \; \left[ v^{(k)}_{\alpha, \text{tr}} \right]^\dagger \cdot M^{(k)}_{\text{grad}} \cdot v^{(k)}_{\alpha, \text{tr}}, 
\end{align}

where the prefactor $f$ is defined as 

\begin{align}
    f(\lambda^{(k)}_{\text{val}}, \tau_k, s_k) &= \frac{1}{1 + \mathrm{exp}\left[ \left( \lambda^{(k)}_{\alpha, \mathrm{val}} + \tau_k \right) / s_k \right]}, 
\end{align}

and the tolerances are set as the Fermi function

\begin{align}
    \tau_k &= \tau \cdot P_{\mathrm{65}} \left[ \left\{ \Delta \lambda^{(k)}_\alpha \right\}  \right], \nonumber \\
    s_k &= s \cdot P_{\mathrm{65}} \left[ \left\{ \Delta \lambda^{(k)}_\alpha \right\}  \right]. 
\end{align}

Here $P_{\mathrm{65}} \left[ \left\{ \Delta \lambda^{(k)}_\alpha \right\}  \right]$ is the 65\% quantile of the distribution of eigenvalue differences $\left\{ \Delta \lambda^{(k)}_\alpha \right\}$ at fixed $k$. $\tau, s$ are fixed $k$-independent parameters characterizing the tolerance to negative eigenvalues. $M^{(k)}_{\text{grad}}$ is the momentum $k$ Gram matrix computed from a smaller sample carrying gradients for backpropagation. For positive eigenvalues $\lambda^{(k)}_{\alpha, \mathrm{val}} \gg - \tau_k$ relatively to the barrier offset, and large batch size $B$ which yields small $s_k$, the prefactor $f \rightarrow 0$; thus, eigenvectors associated with positive eigenvalues do not result in a constraint. On the flip side, negative eigenvalues are penalized with weight $f \rightarrow 1$. 

We note that in general, for gapped models, it suffices to simply set $\tau = s = 1.0$. However, for gapless models studied, we find that we must tune these parameters, and that these parameters depend on the specific model, range and weight of operators used to construct the Gram matrix. This is because the eigenvalues of the true Gram matrix (obtained from the exact expectation values in the true ground state) tends to have extremely small, close to zero eigenvalues. Importantly, we find that one can choose these tolerances by diagonalization of small system sizes (say $L = 16$) and the same values of $\tau, s$ appear to work well at much larger system sizes (at least, up to the $L = 128$ system sizes studied in this work). Moreover, as we discuss in Sec.~\ref{sec:gappedvsgapless}, we show that one can also estimate the value of $\tau$ by examining the distribution of $\lambda / \Delta \lambda$ obtained from sampling the POVM outcome distribution obtained from measurements on the true ground state at system sizes as small as $L = 8$. In some ways, one can view these parameters playing the same role as that of the exchange functional in Kohn-Sham Density Functional theory which requires some degree of knowledge of the underlying model and phase~\cite{Burke2012}. 

We note again that the eigenvectors $v^{(k)}_{\alpha, \mathrm{tr}}$ are obtained from the non-gradient-carrying buffer set and thus determine effectively the weight with which various operators in $M^{(k)}_{\text{grad}}$ are penalized by the constraint objective. 

\subsubsection{Adaptive balancing of loss terms}

The total loss minimized in training is given by
\begin{equation}
\mathcal{L} = \mathcal{L}_E + \lambda_{\mathrm{PSD}} \mathcal{L}_{\mathrm{PSD}}.
\end{equation}

The relative strength $\lambda_{\mathrm{PSD}}$ is not fixed, but determined dynamically by comparing gradient norms:
\begin{equation}
\lambda_{\mathrm{PSD}} = \lambda_{\text{tgt}} \cdot \frac{\|\nabla \mathcal{L}_E\|}{\|\nabla \mathcal{L}_{\mathrm{PSD}}\|}.
\end{equation}

where $\lambda_{\text{tgt}} \sim \mathcal{O} (1)$ ensures that neither term dominates excessively at any point during the training. Over training runs, $\lambda_{\text{tgt}}$ is adapted to ensure that the total violation of the PSD constraints, quantified by 

\begin{align}
    P = \text{max}_{k, \alpha} \left[ f (\lambda^{(k)}_{\alpha, \text{val}}, \tau_k, s_k ) \right]
    \label{eq:P}
\end{align} 

remains finite and small. In practice, we find $\lambda_{\text{tgt}}$ settles at a finite $\mathcal{O} (1) $ value near the end of training.

\subsection{Training schedule and Optimizations}

\subsubsection{Projection of Conflicting Gradients}

We note that the PSD loss objective is not a conventional loss term. It cannot be allowed to grow large as that signals that the optimization routine is exploring unphysical space of states. To stabilize training, it greatly helps to prevent motion in directions that minimize energy but can increase the PSD loss. In particular, we partially project out the gradient due to the energy term $\nabla\mathcal{L}_E$ when these anti-align with the gradient due to the PSD constraint term $\nabla \mathcal{L}_{\mathrm{PSD}}$, 

\begin{align}
    \nabla \mathcal{L}_{E} \rightarrow \nabla \mathcal{L}_{E} - P \; \frac{ \nabla \mathcal{L}_{E} \cdot \nabla \mathcal{L}_{\mathrm{PSD}} }{\left|| \nabla \mathcal{L}_{\mathrm{PSD}} \right||^2}, 
\end{align}

with a strength controlled by the extent of violation of the PSD constrained, quantified again by $P$ defined in Eq.~(\ref{eq:P}). 

\subsubsection{Gumbel-softmax straight-through estimator}

A central difficulty in training autoregressive generative models arises from the discrete nature of the sampled POVM outcomes.  Sampling a discrete outcome from the conditional distribution
\begin{equation}
p(i_j \mid i_{<j})
\end{equation}
is not differentiable, preventing gradients from propagating through the sampling procedure in the standard manner.

To address this issue, we employ the Gumbel-softmax straight-through estimator.  Given logits $\ell_j$ for the conditional distribution at site $j$, we first generate noisy logits
\begin{equation}
\tilde{\ell}_{j,a}
=
\ell_{j,a}
+
g_{j,a},
\end{equation}
where the Gumbel noise variables are
\begin{equation}
g_{j,a}
=
-\log(-\log u_{j,a}),
\qquad
u_{j,a}\sim \mathrm{Uniform}(0,1).
\end{equation}

A differentiable approximation to the sampled one-hot vector is then constructed as
\begin{equation}
y_{j,a}
=
\frac{
\exp\left(\tilde{\ell}_{j,a}/T\right)
}{
\sum_b
\exp\left(\tilde{\ell}_{j,b}/T\right)
},
\end{equation}
where $T$ is a temperature parameter.

In the forward pass, the discrete sample is obtained by taking the hard argmax,
\begin{equation}
i_j = \arg\max_a \tilde{\ell}_{j,a},
\end{equation}
which is then converted into a one-hot vector.  However, during backpropagation, gradients are propagated through the continuous softmax variable $y_j$.  This defines the straight-through estimator.

Operationally, the straight-through procedure allows the model to generate discrete POVM outcomes during sampling while still maintaining differentiability of the computational graph. We typically anneal the temperature $T$ during training, beginning with $T = 1.0$ and gradually approaching nearly discrete sampling limit, $T \rightarrow 0$.  In practice, we find that the results are fairly insensitive to the precise annealing schedule or the value of the final temperature. 

%We note that the use of the straight-through estimator introduces bias in the gradient estimates, since the forward and backward passes correspond to different computational graphs.  Nevertheless, the reduction in gradient variance significantly improves optimization stability, particularly in the presence of the stiff PSD constraints discussed previously.

\subsubsection{Gradient schedule}

The gradient schedule is computed using the AdamW optimizer which we find to be superior for training compared to Adam. We anticipate this is because our optimization already carefully balances multiple competing objectives, and the implicit, gradient-dependent regularization in Adam interferes with that balance in an uncontrolled way. 

The weight decay parameter in AdamW is set to $10^{-3}$, while the beta parameters are annealed from the initial values of $(0.7,0.9)$, which allows rapid changes, to $(0.95,0.99)$ as the training converges. The annealing is governed by the comparison of $\lambda_{\text{tgt}}$ to $\lambda_{\text{tgt,max}} = 1.4$---in general, we find $\lambda_{\text{tgt}}$ grows from a finite initial value of $\lambda_{\text{tgt,min}} = 0.99$ to a $\approx 1.3$ at the training converges. We interpolate between the two sets of beta values noted above linearly with the parameter $x = \text{log}(\lambda_{\text{tgt}}) - \text{log}(\lambda_{\text{min}}) / (\text{log}(\lambda_{\text{tgt,max}}) - \text{log}(\lambda_{\text{min}}) ) $. 

\subsubsection{Optimizations}

The computation of PSD constraints requires evaluating expectations over many operator combinations, leading to significant memory usage. To address this, we employ:

\begin{itemize}
\item Batching of samples: We compute gradients with respect to batches of samples. The overall gradients, particularly for the memory-intensive $\nabla \mathcal{L}_{\text{PSD}}$ can be computed by adding all the batched gradients. 
\item Manual checkpointing: Furthermore, gradients are computed with respect to the single Pauli expectation values from which all possible higher weight Pauli expectation values can be computed by multiplication; see discussion above. Only after such gradients are gathered, is the gradient with respect to model parameters computed. 
\item Operator batching: Gram matrix entries are computed in chunks of operators, not just samples, as the number of operators grows as $L^2$ as system size is increased for finite range $R$.
\item Caching of momentum expansions: We store caching data that helps to rapidly fill the Gram matrices at all momenta $k$.
\end{itemize}

These techniques enable scaling to system sizes up to $L=128$ and beyond while maintaining feasible memory usage. Speed of computations is thus generally constrained by the maximum number of operations that can performed in the GPU cores in parallel, rather than memory usage. 

\vspace{0.1cm}
\section{Numerical Results}
\label{sec:numerics}
\begin{figure}
\centering
    \includegraphics[width=\columnwidth]{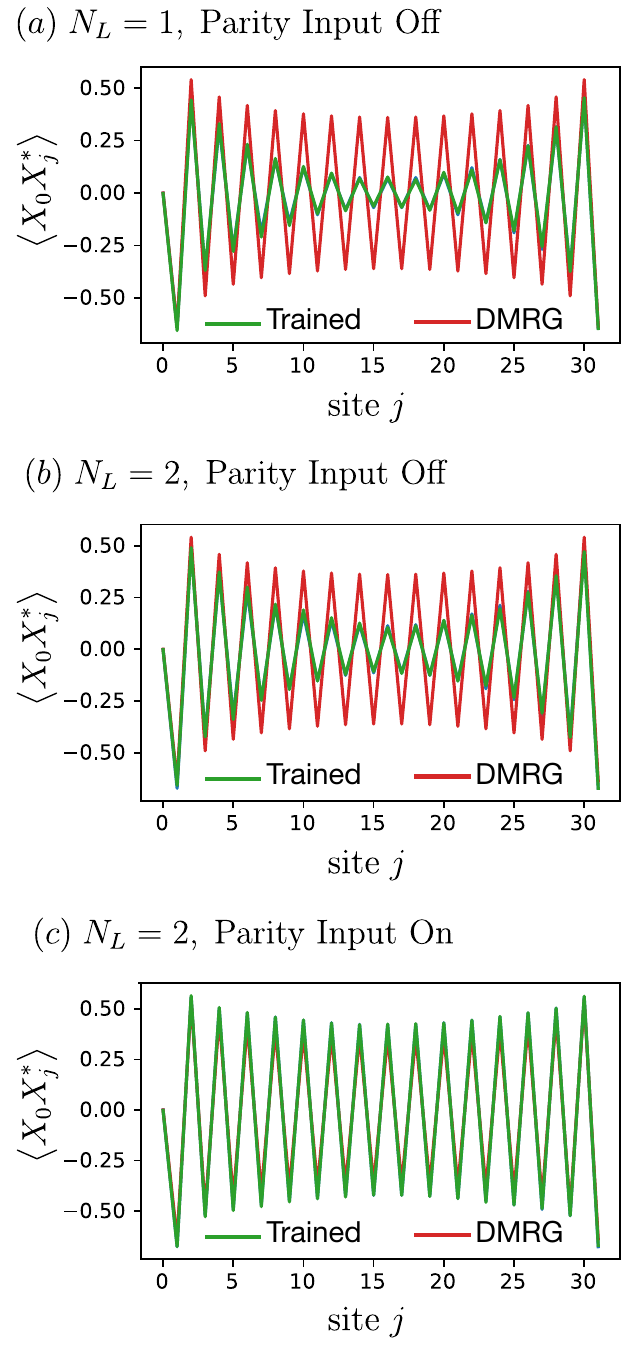}
    \caption{Correlations in the XX channel, $\avg{X_0 X^*_j}$, where $X^*_j = X_j$ for $j \neq 0$, and $X^*_0 = I$, obtained via training (green) and DMRG (red) for the critical TFIM model, $H_{\text{TFIM}}$, with different GRU architectures---standard architecture with no site parity information for (a) $N_L = 1$ and (b) $N_L = 2$ are shown, along with the results for the dual stream architecture in (c) with $N_L = 2$. Adding more layers helps improve correlations but providing parity information is essential in addressing long range correlations. Operator range $R = 4$ is used in obtaining these results.}
    \label{fig:architecture}
\end{figure}

We now present results of training for various system sizes, many-body Hamiltonians, neural network architectures, and underlying parameters. To quantify the success of the training procedure, we will compare the results obtained to those from DMRG (implemented via ITensor~\cite{ITensor}), focusing on the ground state energy density relative to the DMRG computed energy density, $\Delta E/L$, one point $\avg{P_i}$, and two point $\avg{P_i P_j}$ correlations, and the energy variance per site $\Delta H^2/L \equiv \left[ \avg{H^2}-\avg{H}^2 \right]/ L$.

\subsection{Role of Architecture}

While RNNs have been shown to be reasonably adept at learning ground state wavefunctions, it is well known that recurrent and finite-memory autoregressive models can struggle when systems exhibit long-range correlations, such as those in critical models~\cite{yang2024can,ayub2026geometry}. This is because of the autoregressive nature of the update process---the POVM outcome probabilities for the next site, $p_j$, depend on the previous, categorical outcome $i_{j-1} \in \{0,1,2,3\}$, and the hidden state $h_{j-1}$ that carries a summary (and burden) of representing all past measurement outcomes. Naturally, the information pertaining to past outcomes is diluted in the process of successive passes through the RNN. GRUs, which we use, are designed to manage this failure mode better by including learned gates for handling the rate of this memory loss. In conclusion, we can expect the trained state to not only be determined by the training objective, but also the architecture. 

We find the standard GRU-based architecture can be a limiting factor in training.  We illustrate this with the example of training for the ground state of the transverse-field Ising model (TFIM) at criticality. To fix notation, we specifically study the model 

\begin{align}
    H_{\text{TFIM}} = 0.3 \sum_i X_i X_{i+1} + 0.3 \sum_i Z_i, 
\end{align}

with periodic boundary conditions. This has anti-Ferromagnetic correlations $\avg{X_i X_j} \sim (-1)^{\abs{i-j}} f(|i-j|)$, where $f(|i-j|) > 0$ and decays with the characteristic power-law behavior with exponent $1/4$~\cite{giamarchi2003quantum,francesco2012conformal}.  

We show that adopting the dual stream architecture discussed above, which maintains two chains of recurrent neural networks, one which takes as input a one-hot encoding of the previous site measurement outcome, and one which concatenates this input with site parity $(-1)^j$, improves the training outcomes, obtaining correlations that are much closer to the true ground state. The results of trained correlations in the case of a) standard GRU architecture with number of layers, $N_{\rm Layers} = 1$, and b) $N_{\rm Layers} = 2$, and c) dual stream architecture with $N_{\rm Layers} = 2$, are presented in Fig.~\ref{fig:architecture}. 
We have also tested dilated GRU networks in which different layers transmit hidden states at successively longer distances and found minimal improvement over the plain vanilla GRU network with multiple layers, at least for layers up to $N_L = 2$. 

\begin{figure}
\centering
    \includegraphics[width=\columnwidth]{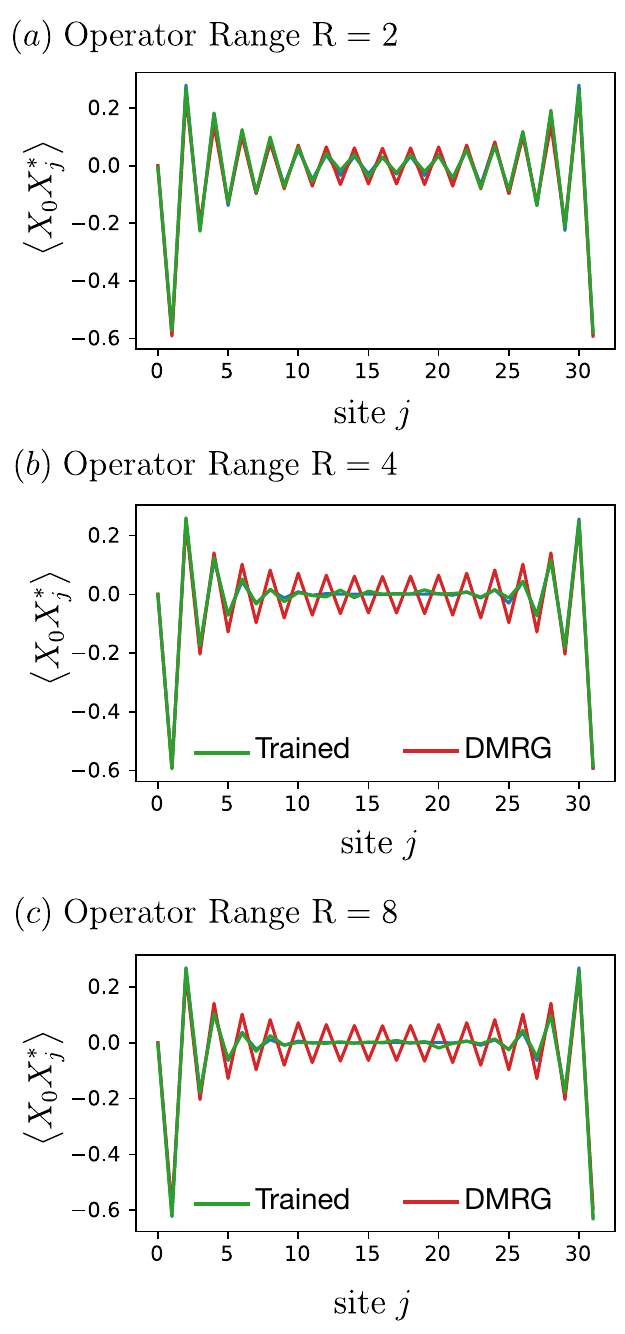}
    \caption{Correlations in the XX channel, $\avg{X_0 X^*_j}$, where $X^*_j = X_j$ for $j \neq 0$, and $X^*_0 = I$, obtained via training (green) and DMRG (red) with different operator ranges (a) $R = 2$, (b) $R = 4$ and (c) $R = 8$. }
    \label{fig:range}
\end{figure}

As we discuss in Sec.~\ref{sec:largerL}, at larger system sizes, it becomes evident that the correlations obtained are likely approximations to power law decay stemming from a combination of finite decay lengths. We anticipate that the dual-stream architecture could benefit from multiple layers that carry momentum-specific information, that is, the site-dependent phase $e^{ik j}$ for relevant momenta $k$, such that each layer can learn a corresponding decay length scale. In this way, the combination of multiple decay scales at different momenta, can then more faithfully approximate the power law. We anticipate that such architectural modifications could be useful beyond the context of this specific work, specifically in wave-function based ansatzes. We have not explored such modifications beyond the inclusion of the $k = \pi$ parity stream as discussed above. Given the relative success of the dual stream approach, subsequently we will exclusively present results obtained by training on the dual stream model. 

\subsection{Dependence on Range and Weight of operators in the Gram matrix}

Another set of important parameters that control the optimization routine is the range and weight of operators that go into the Gram matrix that constrain the obtained POVM outcome distributions to correspond to physical states. In general, one may expect that the larger the pool of operators that enter this matrix, the more accurate the PSD constraint becomes. While this is true technically, it is important to recall that increasing the operator pool also increases the size of the Gram matrix whose eigenvalues we demand to be positive, and concomitantly, the complexity of the potential landscape, as the degree of the polynomial in the outcome probabilities $p_\vs{i}$ is precisely determined by the size of this Gram matrix. Moreover, the entries of the Gram matrix carry stochastic, sampling noise which can also get amplified in the process. 

We find that while increasing the maximum operator weight to at least $w = 2$ is crucial in obtaining reasonable results, increasing the range can in fact deteriorate the optimization procedure. We demonstrate this effect using training on the anti-Ferromagnetic Heisenberg Hamiltonian which is also gapless; see Fig.~\ref{fig:range}. In particular, we optimize with the Hamiltonian 

\begin{align}
    H_{\text{Heisenberg}} = 0.3 \sum_i X_i X_{i+1} + Y_{i} Y_{i+1} + Z_{i} Z_{i+1}, 
\end{align}

with periodic boundary conditions. Similar results can also be observed for $H_{\text{TFIM}}$. We note that in general some of this loss in fidelity as we increase range can be counteracted by increasing the size of the buffer and gradient sets, $B$ and $G$ respectively, which reduce the stochastic noise in determining the eigenvectors corresponding to the most negative eigenvalues, and improve the signal to noise in the gradient computation. Numerically, it is crucial to keep $B, G$ to reasonable numbers, and the range $R$ finite because the number of unique operators in the overall (combining all momenta $k$) Gram matrix scales as $\sim L^2 R^2$. 

\begin{figure}
\centering
    \includegraphics[width=\columnwidth]{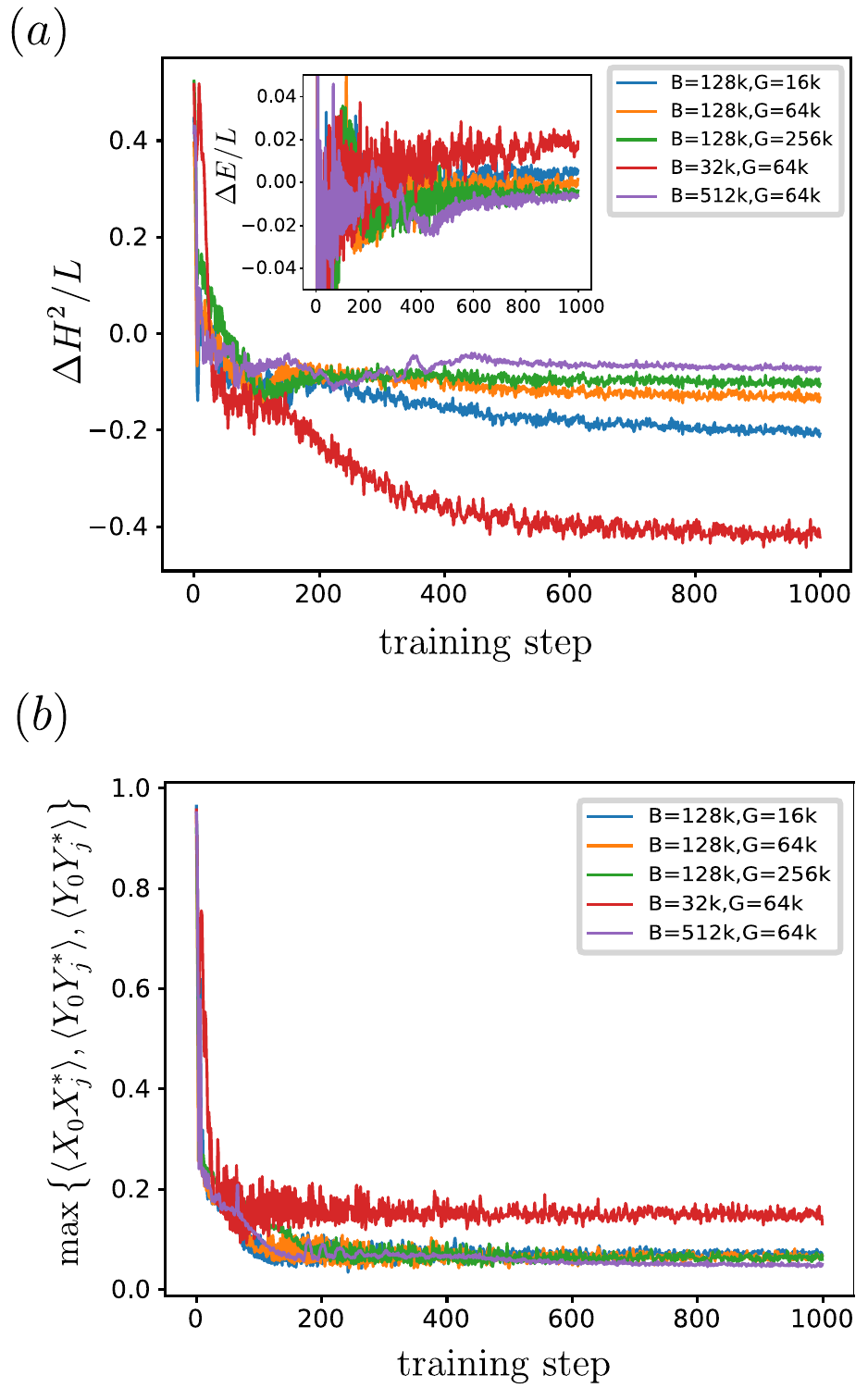}
    \caption{(a) Computation of the energy variance per site, $\Delta H^2 / L$ and (inset) deviation of energy density from DMRG result, $\Delta E/L$ for the TFIM gapped with a $Z$ field, $H_{\text{Gapped TFIM}}$, as a function of the training step, for various buffer and gradient batch sizes, $B$ and $G$, respectively. (b) Maximum deviation in any XX, YY, or ZZ correlators from DMRG results, $\text{max} \{ \avg{X_0 X^*_j}, \avg{Y_0 Y^*_j},\avg{Z_0 Z^*_j} \}$, as a function of training step, for various values of $B, G$. The estimation of these values is obtained using an inference batch size $I = B$; this is generally sufficient to estimate $\Delta E/L$ and the deviation in correlations, but not for $\Delta H^2/L$ which is inferred by summing $\mathcal{O}(L^2)$ operators of weight $w = 4$. Thus, one should not interpret the values of $\Delta H^2/L$ above as the true energy variance in the learned state; see discussion in main text for more details. }
    \label{fig:batchsize}
\end{figure}

It is worth emphasizing that how strongly these Gram matrices constrain the physicality of the correlations and the underlying state overall is not well understood, except that it can be shown that adding to the pool of operators results in tighter constraints and increasing the weight of operators to the system size, $w = L$ guarantees a density matrix that is physical. In previous works on optimizing correlators (as independent optimization parameters and not as in this work where they are obtained from sampling a learned POVM outcome distribution) in reduced density matrices, it was found to be necessary to include all (arbitrary range) two-body operators to obtain reasonable fidelity with the true ground state. In the present instance, as noted above, we start with a learned POVM distribution from which these correlators are computed by inverting the shadow channel (or utilizing the dual operators). Thus, additional constraints are already built into the inference procedure because one cannot obtain completely arbitrary values for the expectation values of various operators from the POVM outcome distribution. Thus, it is plausible that a smaller set of weight $w = 2$ operators with maximum range $R = 2$ already provides reasonably-strong constraints. Our results suggest this to be the case. 

\subsection{Dependence on increased buffer and gradient batch sizes}

In general, the results improve with increasing buffer and gradient batch sizes. This is to be expected if the optimization behavior is dominated by the stochasticity in the inference of correlators from learned POVM outcome distributions. (Note that this is not guaranteed to be the case because our state is constrained by partial set of constraints determined by the weight and range of the operator pool in the Gram matrices, which can allow for unphysical underlying state and concomitantly, deviations from exact results.) 

We demonstrate the improvement with batch sizes by running the optimization procedure on a gapped TFIM model, given by the Hamiltonian 

\begin{align}
    H_{\text{Gapped TFIM}} = 0.3 \sum_i X_i X_{i+1} + 0.6 \sum_i Z_i
\end{align}

Note that the variance can be negative in this framework because the inferred density matrices are not entirely physical; likewise the energy can be below the true ground state energy. In practice, we find the energy density to be around $0.01$ off from DMRG results, and the maximum error in any weights $1$ and $2$ correlations to be around $0.05$, for the larger batch sizes used. This appears to be relatively consistent across the models we have studied. From Fig.~\ref{fig:batchsize} (a), we see that increasing the buffer size by a factor of $4$ tends to reduce the energy variance per site, $\Delta H^2 / L$ by a factor of $2$; this behavior also applies approximately for the maximum error in all weight $1$ and weight $2$ correlations, as seen in Fig.~\ref{fig:batchsize} (b). This appears consistent with the scaling expected from the variance of operators inferred from finite number of samples drawn from the POVM outcome distribution; Eq.~(\ref{eq:var_w}).

\begin{figure}
\centering
    \includegraphics[width=0.9\columnwidth]{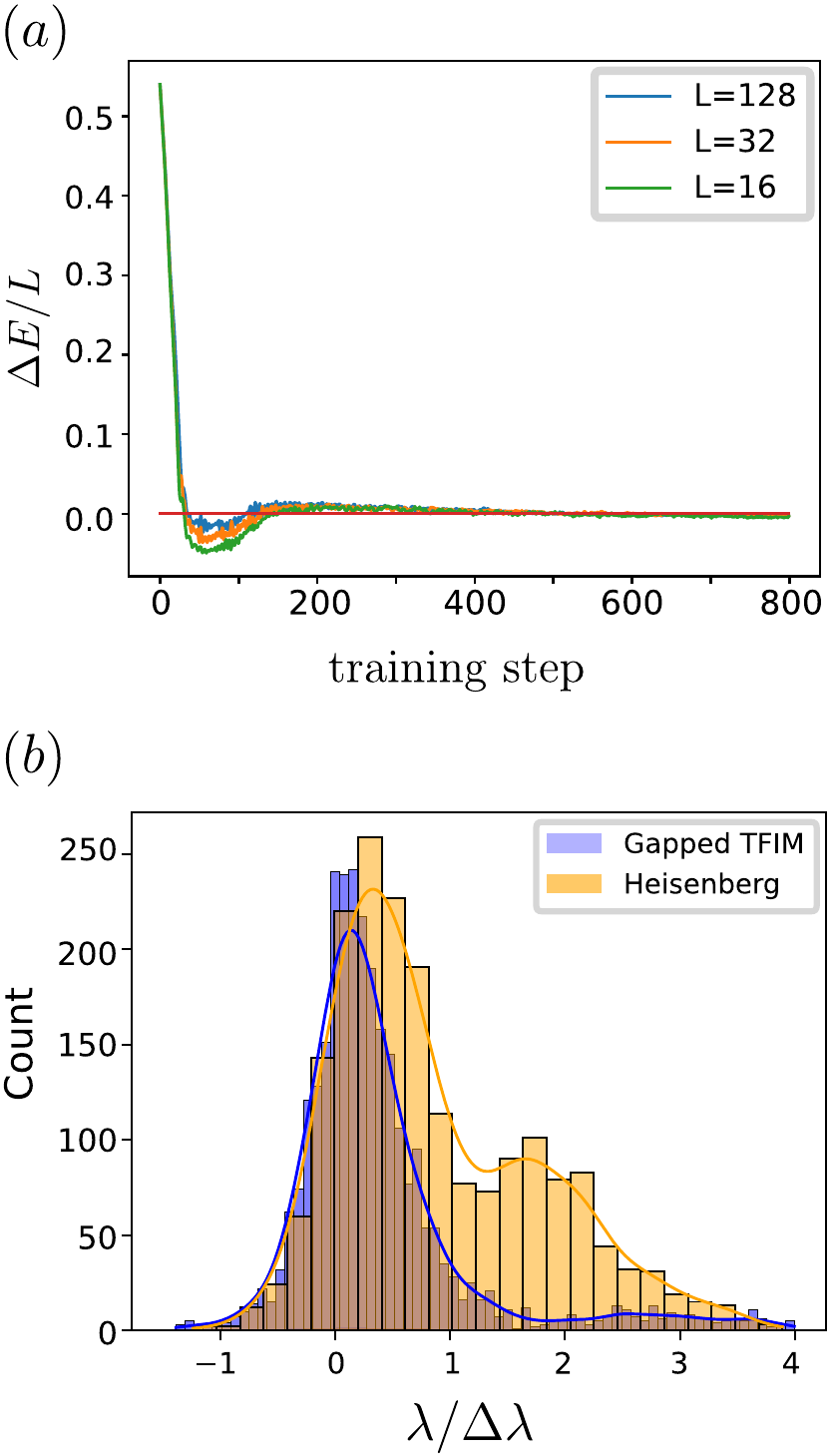}
    \caption{(a) Deivation of energy density for $H_{\text{Heisenberg}}$ obtained from training to that obtained by DMRG for $L = 16, 32, 128$, and fixed scaling factor $\tau = 4.0$ that determines tolerance of negative eigenvalues of the Gram matrix, indicating that it can be fixed based on smaller system size simulations. Here we employed a buffer batch size of $128k$ samples. (b) Comparison of normalized Gram matrix eigenvalues $\lambda$ obtained by sampling POVM outcomes from the ground states of  $H_{\text{Gapped TFIM}}, H_{\text{Heisenberg}}$ obtained by exact diagonalization at $L=8$. $\Delta \lambda$ is obtained analogously to the training algorithm, with the samples divided into two halves. Gram matrices constructed from these two sample sets differ due to sampling noise and $\Delta \lambda$ characterizes the typical difference in eigenvalues; see main text for more details. The set of small eigenvalues for the Heisenberg model is broader than that of the gapped model which is approximately Gaussian with width $1$.}
    \label{fig:GappedGapless}
\end{figure}

\begin{figure}[h]
\centering
    \includegraphics[width=0.9\columnwidth]{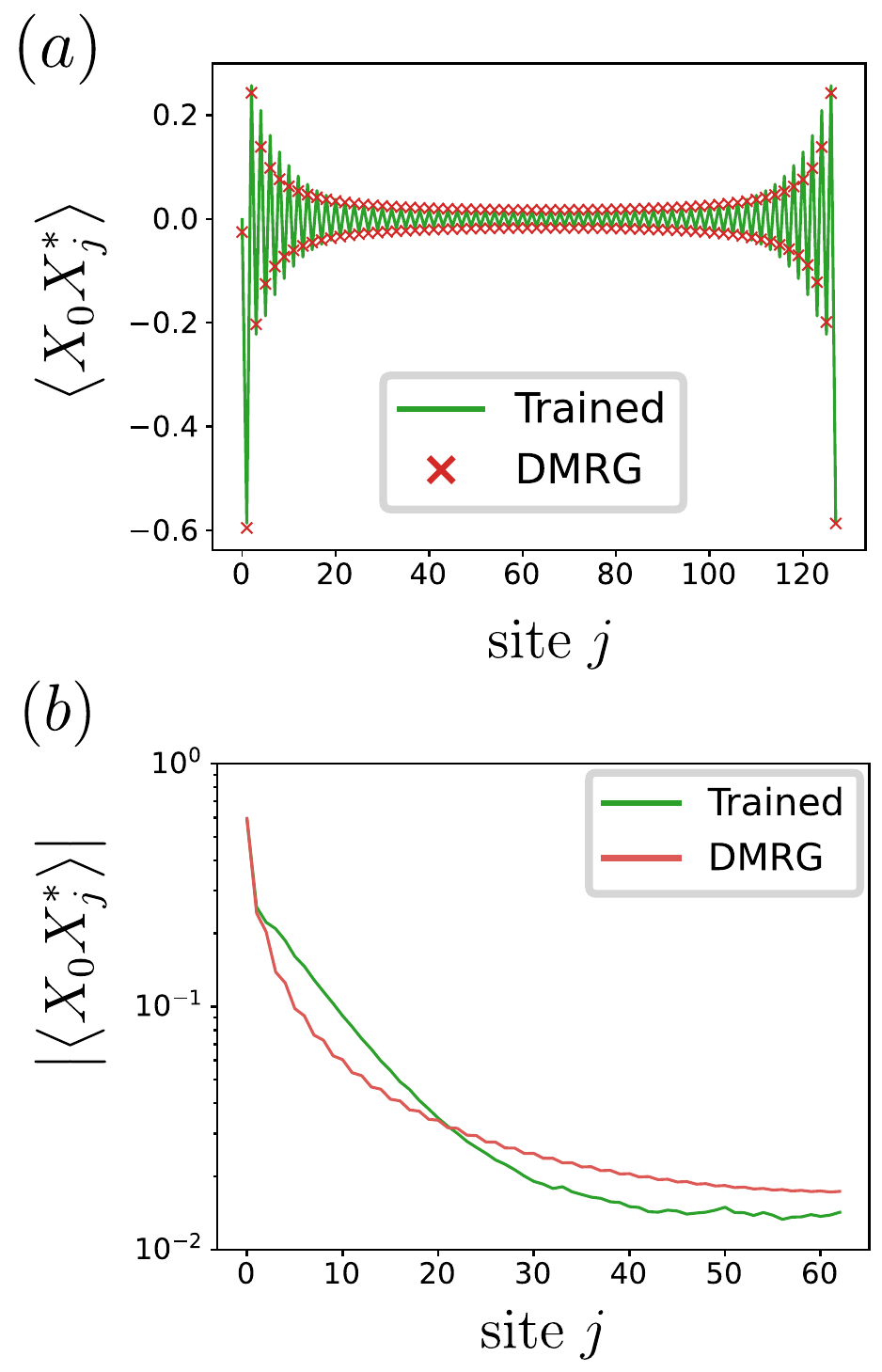}
    \caption{Correlations in the XX channel, $\avg{X_0 X^*_j}$, where $X^*_j = X_j$ for $j \neq 0$, and $X^*_0 = I$, obtained via training (green) and DMRG (red) on (a) linear and (b) log scale, for $H_{\text{Heisenberg}}$ and $L = 128$. We see several points of discrepancy. At site zero, where we $X^*_0 \equiv \mathbb{I}$, we see that the DMRG computation results in a non-zero expectation value of local spin, while the training routine results in a value close to zero, which is the theoretically expected result. This is likely due to the DMRG being challenged by periodic boundary conditions, particularly, at $L = 128$. We see that the algorithm gets the long-range value of the correlators correct but at the expense of shorter range correlators; this is particularly evident for sites $j = 2,3,..,10$. On the log scale, the reason for this deviation becomes more apparent---the training routine is attempting to approximate the critical power law behavior with an exponentially decay. This likely stems from the autoregressive nature of the model which naturally embodies a decay length, and could be addressed by a transformer-based network with long range attention.}
    \label{fig:L128Heis}
\end{figure}
We note here that it is important to distinguish errors coming from finite sampling variance, as per Eq.~(\ref{eq:var_w}), as opposed to errors originating from deviations of the learned POVM outcome distribution from the  outcome distribution obtained in the true ground state. Here we have presented all results with the same batch size for inference as the buffer batch size $B$ used in training. We note that for $B = 128000$, one can estimate the sampling variance of any weight $2$ operator to be $\sqrt{\text{Var}[O_{w=2}]} \approx 0.025$, which is comparable but nevertheless smaller than the amplitude of maximum deviation seen between DMRG results and trained weight $2$ correlators, which is of the order of $\approx 0.05$. Likewise, the trained energy density, which is composed of $2$ terms per site for $H_{\text{Gapped TFIM}}$, deviates from the DMRG energy density by about $\approx 0.01$. (If the terms in the energy computation were statistically independent, we would obtain a further reduction by $1/\sqrt{L}$ in sampling noise, but since the model is translationally invariant, the terms are likely closer to being perfectly correlated.) Thus, we anticipate that the errors obtained for the energy density and the correlators are likely dominated by a combination of sampling noise, but also true deviations in the learned POVM outcome distribution which arise due to training with finite buffer batch sizes. 

%Similarly, it is important to note that the value of $\Delta H^2 / L$ is obtained as a sum of $\mathcal{O} (L^2)$ operators which are predominantly weight $w = 4$. 

%This finite sampling noise in inference is unlikely to be a factor in the estimation of any two-point correlator given that $\sqrt{\text{Var}[O_{w=2}]} \approx 0.025$ for batch size $128000$ which is below the max deviation $\approx 0.06$ in the correlator values obtained from training in comparison to DMRG results. 

\subsection{Gapped vs Gapless Hamiltonians}
\label{sec:gappedvsgapless}

In general, given the absence of an energy gap, one expects that finding the ground state of gapless Hamiltonians should be a more challenging task than finding the ground state of gapped Hamiltonians. Indeed, there are certain aspects of this that hold true in these results---it is generally harder to learn non-trivial longer range correlators that show up in gapless systems. However, the  number of training steps needed for convergence do not appear to be markedly different. 

However, we find that the treatment of constraints is more subtle in the case of gapless models. In particular, for the gapped models we have tested, such as the Heinseberg model gapped by a uniform field, or the transverse-field Ising model with an applied field as discussed above, we find that the parameters $\tau, s$, which set the barrier offset and the barrier width, respectively, can be simply set to $1.0$ to obtain optimal results. On the other hand, for gapless Hamiltonians, we find that convergence to a state with high fidelity to ground state energy and correlations obtained from DMRG requires tuning the parameters $\tau, s$ away from $1.0$. 

This is somewhat unsatisfactory as we expect to be able to obtain the ground state without resorting to ad-hoc tuning of any parameter. To this end, we note that the choice of $\tau, s$ that yield optimal results seem to be independent of system size, and we may set it based on results at small system sizes where accurate data is available; we provide evidence of this in Fig.~\ref{fig:GappedGapless} (a). 

Furthermore, we find that one can obtain a reasonable estimate of the optimal choice of $\tau$ by examining the distribution of the eigenvalues of the Gram matrices by sampling the true ground state at very small system sizes. For instance, to obtain accurate results for the Heisenberg model, we have used the values $s = 1.0, \tau = 4.0$ for operator weight and range $w = 2, R = 2$. In Fig.~\ref{fig:GappedGapless} (b), we show the eigenvalue distribution obtained by generating POVM outcome samples by performing measurements on the true ground states of the Heisenberg model, $H_{\text{Heisenberg}}$ and the gapped Ising model $H_{\text{Gapped TFIM}}$, for $L = 8$. (Note that the exact POVM outcome distribution is already a vector of probabilities of $4^8 = 65,536$ unique measurement outcomes.) Subsequently, the Gram matrices are constructed from these samples and diagonalized to obtain eigenvalues $\lambda$. These eigenvalues are normalized by $\Delta \lambda$, which is obtained precisely as a the $65^\text{th}$ quantile of the distribution of eigenvalue differences between a `training' and `validation' set of samples as discussed in Sec.~\ref{sec:architecture}. Finite values of $\lambda / \Delta \lambda$ should be seen as $\approx 0$ eigenvalues of the Gram matrix, up to sampling noise. We see that the distribution of $\lambda / \Delta \lambda$ in the case of gapped TFIM model is essentially uniform about $0$ with a width $\lambda/ \Delta \lambda = 1.0$. On the other hand, the distribution of eigenvalues is the Heisenberg model extends on the positive side to around $\lambda / \Delta \lambda = 4.0$. We interpret this as the statement that the ground state of the Heisenberg model has a lot more eigenvalues close to $0$ than the gapped TFIM model, and the factor of $\tau = 4.0$ used to obtain accurate results appears to coincide with the width of the distribution found from measurements on the exact ground state at $L = 8$.

%In this approach, we do not find considerable difference between gapped and gapless models---the training steps needed for convergence appear to be relatively similar. We note however that the accuracy of the results may not be sufficient anyway to distinguish between the ground state and other low-energy eigenstates. 

\subsection{Scaling to larger system sizes}
\label{sec:largerL}

The rate-limiting step of this approach is the computation of all operators that enter into the Gram matrix over all momenta. As noted above, if the operator pool consists of finite maximum weight $w$ operators with finite range $R$, the linear dimension of the Gram matrix scales as $\sim R^{w-1} \cdot L$. Populating this Gram matrix thus requires $\mathcal{O} (L^2 R^{2(w-1)}) $ operations. For $w=2$ considered in this work, almost all other steps, including the generation of the samples from the underlying probability distribution, backpropagation to compute gradients with respect to model parameters, diagonalization of fixed momentum Gram matrices, are done efficiently and require a negligible amount of time in comparison. In total, the overall scaling with system size is highly favorable, and since expectation values of all observables is a simple average of contributions from each sample, the procedure is trivially parallelizable over multiple GPUs. 

In Fig.~\ref{fig:L128Heis}, we show the results of the procedure as applied on the critical Heisenberg model, $H_{\text{Heisenberg}}$, for $L = 128$, range $R =2$, which shows reasonably good agreement with DMRG results. Note that for such system sizes, DMRG itself has difficulty dealing with the gapless nature of the system, particularly with periodic boundary conditions, and produces erroneous finite expectation values of observables such as local spin polarization where they must be zero, while the present approach gets this aspect right. However, plotting the findings on a log scale, it becomes clear that the training routine essentially tries to approximate the critical power-law decay with an exponential form. While one can then obtain reasonably accurate long-range behavior, the short range correlations end up deviating appreciably from the DMRG results. We believe this is largely due to the autoregressive architecture employed and not a shortcoming of the training objective and constraints themselves. While GRUs and the dual stream approach can help in memory transmission over long range, helping these correlations track exact (DMRG) values, it fundamentally cannot alter the fact that the architecture is biased towards states that have a finite length scale that governs decay; see for instance the discussion in Ref.~\cite{}. A transformer based architecture with attention should be able to resolve this issue but we have not explored this in the present work. 

\section{Conclusions and Outlook}
\label{sec:conclusions}

In this work, we explored variational optimization of quantum many-body states directly in informationally complete POVM space using autoregressive recurrent neural networks. Rather than representing wavefunction amplitudes, the neural network learns a probability distribution over POVM outcomes, from which observables and density matrices are reconstructed through the dual operator frame.

This formulation leads to an optimization problem whose structure differs qualitatively from conventional neural quantum states. In particular, the energy objective is linear in the learned probabilities, while the dominant geometry of the optimization landscape arises from the positivity constraints required to ensure that the reconstructed density matrix is physical. We implemented these constraints through momentum-resolved operator Gram matrices constructed from low-weight Pauli operators and studied their interaction with autoregressive neural-network architectures.

Several observations emerge from this study. First, the choice of neural-network architecture strongly affects the ability to reproduce long-range critical correlations. In particular, standard GRU-based autoregressive models struggle to faithfully represent staggered antiferromagnetic correlations. Introducing physically motivated inductive biases through parity-aware dual-stream recurrent architectures substantially improves the learned correlations in critical systems. This suggests that autoregressive neural quantum states may benefit significantly from incorporating momentum-selective or symmetry-aware processing pathways. We note that despite obtaining better results with the dual-stream architecture, the model still appears to approximate genuine power-laws in critical models with a combination of finite length scales; using dilated recurrent neural networks~\cite{ayub2026geometry}, or incorporating multipler layers focussed on different momenta, may alleviate this issue. 

Second, we find that enlarging the operator pool used in the positivity constraints does not necessarily improve optimization. While increasing operator range and constraint complexity formally strengthens the physicality conditions, it also amplifies stochastic fluctuations and complicates the optimization landscape. In practice, relatively local constraints often yield the best overall results. This observation suggests that the effectiveness of positivity constraints is controlled not only by their formal strength but also by their compatibility with finite-sampling optimization. At the same time, our results show that a fairly limited set of operators, with finite range and finite weight, does provide strong enough constraints to approach close to the ground state of a wide variety of models. This also facilitates a very favorable scaling of the computational cost of the algorithm  $\sim \mathcal{O} \left ( L^2 \right) $ in system size $L$.  

Third, finite-sampling fluctuations play an important role in training. We found it necessary to introduce momentum-dependent tolerances calibrated directly from stochastic eigenvalue fluctuations of the Gram matrices. In gapless systems, these tolerances become especially important due to the presence of near-zero eigenmodes in the exact physical Gram matrices. 
%The resulting optimization problem therefore resembles a constrained feasibility problem in the presence of stochastic uncertainty, rather than a conventional smooth variational optimization.

%Despite these advances, the present approach also exhibits important limitations. While local observables and energies can often be reproduced accurately, obtaining quantitatively correct long-range critical correlations remains difficult. In particular, we find evidence that the recurrent autoregressive architecture itself may limit the representation of critical long-distance structure, even when local correlations and positivity constraints are well satisfied. Increasing recurrent depth or introducing dilated recurrent connections does not substantially alleviate this issue.

%More broadly, our results suggest that optimization in POVM space may possess fundamentally different failure modes than wavefunction-based variational approaches. The positivity constraints introduce highly nontrivial optimization geometry, while the autoregressive representation may favor states with effectively finite correlation lengths. Understanding the interplay between these effects remains an important open problem.

The POVM framework possesses several attractive conceptual features. The representation is naturally formulated in terms of experimentally accessible quantities, and applies equally well to mixed states. Furthermore, the positivity constraints derived from operator Gram matrices potentially provide a flexible and physically transparent framework for setting the biases of the trained network and learned probability distributions. At the same time, the accuracy of obtained results suggests that improvements in the architecture---going from the dual stream approach to multiple layers tackling different momenta, or moving away from an autoregressive sampling to global sampling---are needed, as well as developing a clearer understanding of how various constraints limit the kinds of states realized. 

We note that for training, we did not employ stochastic reconfiguration techniques, which generally are needed to accelerate training in the wavefunction-based methods, and which increase the complexity of the algorithm to $\mathcal{O} (N_p^3)$ where $N_p$ is the number of parameters, making scaling to larger neural networks challenging (though certain methods like K-FAC~\cite{Psiformer2023} or MinSR~\cite{Rende2024,Chen2024} provide relief). It is possible that similar ideas would also accelerate training in the POVM space though we have not implemented this. Given the reality and positivity of the POVM outcome probability distribution in this framework, it is also possible that associated geometric tensor is better conditioned than the corresponding one obtained from computing derivatives of complex wavefunctions. Moreover, there is no analog in this setting of the redundancy of global phase of the wavefunction, a further simplification. This suggests that dynamics, wherein the geometric tensor enters naturally, may be simpler, and more robust in the POVM framework.   

Several future directions appear promising. Transformer-based architectures or diffusion models may provide improved long-range expressivity compared to recurrent networks~\cite{Psiformer2023,Viteritti2023,Rende2024,viteritti2025transformer,Chen2024}. More sophisticated positivity constraints, potentially incorporating adaptive operator pools or learned operator bases, may improve physicality enforcement while reducing optimization stiffness. Extending the framework to dissipative dynamics and finite-temperature states is also very promising direction, for reasons noted above. 

%Finally, beyond the immediate numerical performance of the method, we hope this work clarifies some of the conceptual and practical challenges associated with variational optimization directly in measurement space. In particular, our results suggest that while POVM-based representations provide a compelling and physically natural framework, the associated optimization problem is subtle and strongly shaped by the structure of the imposed physicality constraints.

\section{Acknowledgements and Data availability}

We thank Pierre-Gabriel Rozon for useful comments and collaboration on previous related work. The author acknowledges funding from Department of Energy, Office of Science, Basic Energy Sciences for this work. We also acknowledge support from the Argonne Leadership Computing Facility (ALCF) whose clusters were employed to obtain all the results. Code related to this project can be obtained from the GitHub repository \url{https://github.com/kartiek27/ShadowLearning_GAN/tree/different-SIC-capability}.   

\bibliography{refs_expanded2}

\end{document}